\documentclass[sigconf]{acmart}

\AtBeginDocument{%
  \providecommand\BibTeX{{%
    \normalfont B\kern-0.5em{\scshape i\kern-0.25em b}\kern-0.8em\TeX}}}

\copyrightyear{2020}
\acmYear{2020}
\setcopyright{acmcopyright}\acmConference[ICPP '20]{49th International Conference on Parallel Processing - ICPP}{August 17--20, 2020}{Edmonton, AB, Canada}
\acmBooktitle{49th International Conference on Parallel Processing - ICPP (ICPP '20), August 17--20, 2020, Edmonton, AB, Canada}
\acmPrice{15.00}
\acmDOI{10.1145/3404397.3404404}
\acmISBN{978-1-4503-8816-0/20/08}

\acmSubmissionID{pap132}

\usepackage[ruled,linesnumbered]{algorithm2e}
\usepackage{graphicx}
\usepackage{subcaption}
\usepackage{booktabs}
\usepackage{multirow}
\usepackage{adjustbox}
\usepackage{tikz}
\usepackage{pgfplots}
\pgfplotsset{compat=1.14}
\usepgfplotslibrary{fillbetween}
\usepackage{xcolor}
\usepackage{soul}
\usepackage{kotex}

\begin{document}


\title{ShadowTutor: Distributed Partial Distillation for Mobile Video DNN Inference}

\author{Jae-Won Chung}
\affiliation{
  \institution{Seoul National University}
  \city{Seoul}
  \country{South Korea}}
  \authornote{corresponding authors}
\email{jaywonchung@snu.ac.kr}

\author{Jae-Yun Kim}
\affiliation{
  \institution{Seoul National University}
  \city{Seoul}
  \country{South Korea}}
\email{jaeykim@altair.snu.ac.kr}

\author{Soo-Mook Moon}
\affiliation{
  \institution{Seoul National University}
  \city{Seoul}
  \country{South Korea}}
  \authornotemark[1]
\email{smoon@snu.ac.kr}


\begin{abstract}

  Following the recent success of deep neural networks (DNN) on video computer vision tasks, performing DNN inferences on videos that originate from mobile devices has gained practical significance.
  As such, previous approaches developed methods to offload DNN inference computations for images to cloud servers to manage the resource constraints of mobile devices.
  However, when it comes to video data, communicating information of every frame consumes excessive network bandwidth and renders the entire system susceptible to adverse network conditions such as congestion.
  Thus, in this work, we seek to exploit the \emph{temporal coherence} between nearby frames of a video stream to mitigate network pressure.
  That is, we propose \textit{ShadowTutor}, a distributed video DNN inference framework that reduces the number of network transmissions through intermittent \textit{knowledge distillation} to a student model.
  Moreover, we update only a subset of the student's parameters, which we call \textit{partial distillation}, to reduce the data size of each network transmission.
  Specifically, the server runs a large and general \textit{teacher} model, and the mobile device only runs an extremely small but specialized \textit{student} model.
  On sparsely selected \emph{key frames}, the server partially trains the student model by targeting the teacher's response and sends the updated part to the mobile device.
  We investigate the effectiveness of ShadowTutor with HD video semantic segmentation. 
  Evaluations show that network data transfer is reduced by 95\% on average.
  Moreover, the throughput of the system is improved by over three times and shows robustness to changes in network bandwidth.
\end{abstract}




\maketitle

\section{Introduction}
Deep learning approaches have proved to be extremely powerful in fields such as computer vision \cite{MaskRCNN}, natural language processing \cite{Transformer}, speech recognition \cite{LAS}, and sequential recommendation \cite{GRU4Rec}.
Due to their applicability to various services, such DNN models sparked interest for performing DNN inference directly on data generated on mobile devices.

Video is one of the most important among such data due to its numerous practical applications.
Autonomous vehicles perform road segmentation to traverse the street and detect obstacles \cite{Autonomous}.
CCTV cameras used for home security can notice movements and detect objects using DNNs \cite{CCTV}.
Mobile selfie apps segment humans and the rest to change the background or simulate background blurring effects \cite{GoogleFilter}.
Thus, video DNN inferences on comparatively weak devices have become a core mechanism for delivering various services and technologies.




While running a large model directly on-device may yield high accuracy, execution latency and memory consumption can quickly become unacceptable for resource-constrained mobile devices.
Thus, to lift the mobile device of the unwieldy inference workload, several approaches offload DNN computations to cloud servers \cite{NeuroSurgeon,MCDNN,EPOffCloud}.
In this scheme, the mobile device sends video frames to the cloud server and retrieves inference results.
However, these approaches are limited in that information of every frame must be communicated between the server and the client.
This heavily pressures network traffic, which is a view shared with \cite{MSVideoAnalytics}.
Moreover, the entire system becomes vulnerable to adverse network conditions such as reduction in available network bandwidth due to congestion.



To this end, we propose \textit{ShadowTutor}, a general video DNN inference framework.
Here, a very small \emph{pre-trained} student model runs on a mobile device for test time inference.
However, due to its size, the student lacks generalization power to excel on any given input scene even if it has undergone "public education".
Thus, to enhance performance specifically on the video at hand, the student receives "shadow education" from a large teacher model on the server.
Especially, the student receives training only on a fraction of all frames in the video, which frames we call \textit{key frames}.

The student model on the mobile device processes video frames one by one in an online manner.
On sparse \textit{key frames}, the mobile device sends the frame to the server.
Upon key frame receive, the server updates a copy of the student by considering the teacher's inference result as the key frame's label, and returns the updated student weights to the mobile device.
In effect, the student is specialized to the current video stream by distilling knowledge from the teacher during test time.
Now, the student model can accurately perform inference on-device without the teacher's help for \textit{non-key frames}, thanks to the \emph{temporal coherence} between the key frame and the non-key frames after it.
That is, non-key frames are likely to share characteristics such as background, ambience, overall texture, speed of movement, and the objects present with a nearby key frame, thus allowing the student trained on that key frame to accurately process non-key frames that come after it.
Moreover, the mobile device does not wait until the updated student weights arrive; it just proceeds to the next non-key frame \emph{asynchronously} with the slightly outdated student, thereby exploiting temporal coherence further.
Finally, the stride to the next key frame is determined by a simple but adaptive algorithm based on the student's performance after distillation.

Sending the updated weights to the mobile device incurs little overhead because the student is very small by design.
For instance, in our experiment, the student model is even smaller than a single video frame.
Still, we propose \textit{partial distillation}, a technique that further reduces the transmitted data size, and apply it to ShadowTutor.
That is, upon distilling knowledge from the teacher, only a subset of the student's parameters is trained.
This technique accelerates knowledge distillation and allows only the updated part of the student to be transmitted across the network.
Moreover, surprisingly, partial distillation yields generally higher accuracy and further reduces the number of network communications compared with adapting all parameters.


We evaluate the effectiveness of ShadowTutor on HD video semantic segmentation, a challenging video computer vision task that requires dense predictions of pixel-level class probabilities.
Evaluations show that compared with naive DNN offloading, ShadowTutor reduces network data transfer by 95\% on average and improves throughput by over 3x.
Also, due to asynchronous inference, ShadowTutor effectively retains throughput under low available network bandwidth.


In sum, we make the following contributions:

    
    
    

\begin{itemize}
    \item We design ShadowTutor, a distributed video DNN inference framework that reduces the number of network transmissions and the data size of each.
    
    \item We allow ShadowTutor to be robust to changes in network conditions by adopting asynchronous inference.
    
    \item We aid the configuration of ShadowTutor by providing analytic models of its network traffic and throughput in terms of system parameters and component measurements.
    
    \item We adapt ShadowTutor to HD video semantic segmentation, a real-life application, and evaluate its performance in terms of throughput, network traffic, accuracy, and robustness.
\end{itemize}

The rest of this paper is organized as follows.
We review previous approaches in section \ref{sec:previous-approaches}, and provide background regarding our work in section \ref{sec:backgrounds}.
We explain the core idea and mechanism of ShadowTutor in section \ref{sec:ShadowTutor}, and show how it is adapted specifically to video semantic segmentation in section \ref{sec:ShadowTutor-semseg}.
Then, we evaluate our framework in section \ref{sec:evaluation}.
Finally, we introduce related work in section \ref{sec:related-work} and conclude the paper in section \ref{sec:conclusion}.

\section{Previous Approaches} \label{sec:previous-approaches}


Due to the resource constraints of mobile devices, DNN computations are often offloaded to central cloud servers.
Neurosurgeon \cite{NeuroSurgeon} partitions a given DNN and collaboratively executes it between the cloud and the server.
It searches for an optimal partition point in terms of latency and power consumption.
Eshratifar et al \cite{EPOffCloud} also seeks to partition DNNs, but by deriving optimal conditions for client power consumption and cloud resource constraints.
MCDNN \cite{MCDNN} generates modified and specialized versions of the original model to trade-off accuracy for improved resource usage.
However, it only deals with models that perform image classification, requires the execution of the model for every image in the training set for specialization (pre-forwarding), and the generated model is mostly limited to being a layer-wise modification of the original model.
Moreover, adopting these approaches still require the communication of every frame across the network, thereby pressuring network traffic greatly.

Another work by Yang et al. \cite{FlowDist} splits video frames into partitions that are distributed to slaves.
It utilizes optical flow, i.e. the movement vectors between two frames, to exploit the temporal coherence between frames.
However, their system still requires the communication of every single frame between the master and the slaves, since they utilize temporal coherence only to occasionally reduce the amount of computation, rather than reducing network traffic.

\section{Background} \label{sec:backgrounds}


\subsection{Knowledge Distillation}

\begin{figure}[ht]
  \includegraphics[width=7.5cm]{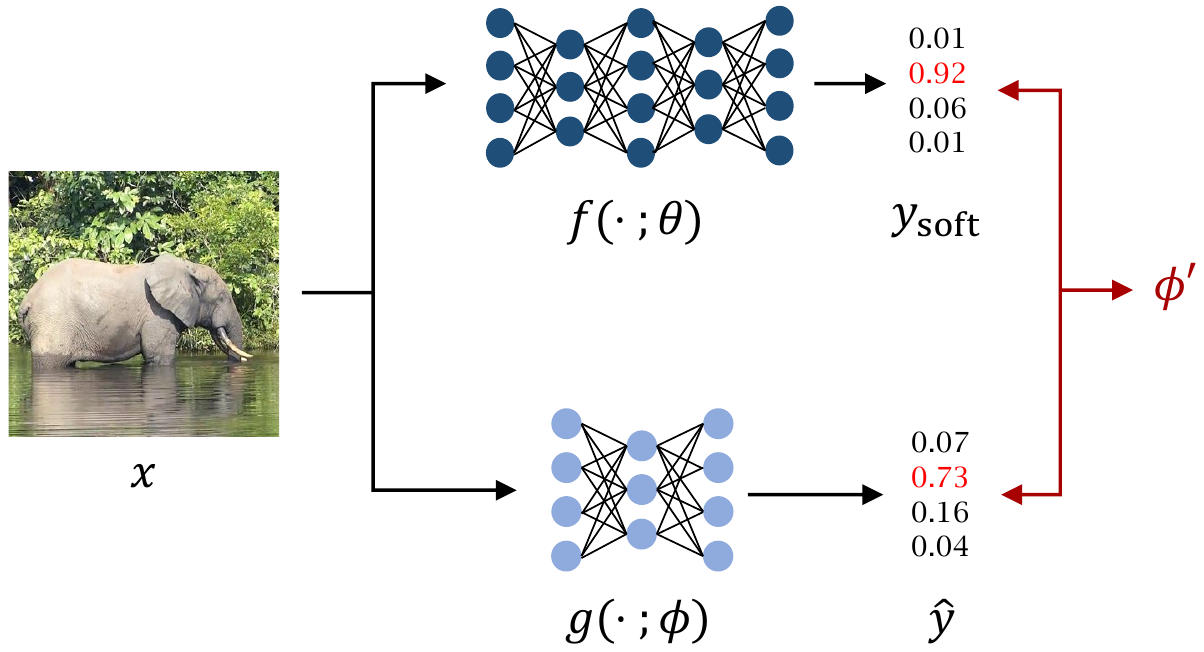}
  
  \caption{Inference time knowledge distillation. The student's parameters are updated by viewing the teacher's ouptut as the label.}
  \label{fig:KD}
\end{figure}

Knowledge distillation \cite{KD} was initially proposed as a means of model compression.
Viewing a large teacher model's output as a \emph{soft target}, it formulated a knowledge distillation loss between the output of a small student model and the soft target.
Compared with \emph{hard targets}, which are one-hot vectors from the dataset, soft targets provide much more information about the input data, and allow less variance in the gradients of the student model.

Figure \ref{fig:KD} illustrates the concept of knowledge distillation applied to image classification.
Since it assumes inference time, the actual label $y_\text{hard}$ is not available.
Thus, the student learns just from the output of the teacher model $y_\text{soft}$.
Is is also the case for ShadowTutor.

\subsection{Semantic Segmentation}

Semantic segmentation is in other words \textit{pixel-wise classification}.
There is a fixed set of classes, including the background class.
Then, the model is required to classify each pixel of its input image to one of the given classes.
Thus, the output of semantic segmentation has the same spatial size as the input image.

The result of semantic segmentation is often evaluated with mean IoU (Intersection over Union).
With $pred_c$ the set of pixels classified as class $c$ and $label_c$ the set of pixels with ground truth class $c$, the IoU for class $c$ is defined as
\begin{equation}
    \text{IoU}_c = \frac{|pred_c \cap label_c|}{|pred_c \cup label_c|}.
\end{equation}
The IoU is computed for each class in the ground truth label and averaged to yield the mean IoU performance metric.
Mean IoU thus lies between 0 and 1, with 1 being the best possible score.

\section{ShadowTutor} \label{sec:ShadowTutor}

\begin{figure*}[t]
  \centering
  \includegraphics[width=\textwidth]{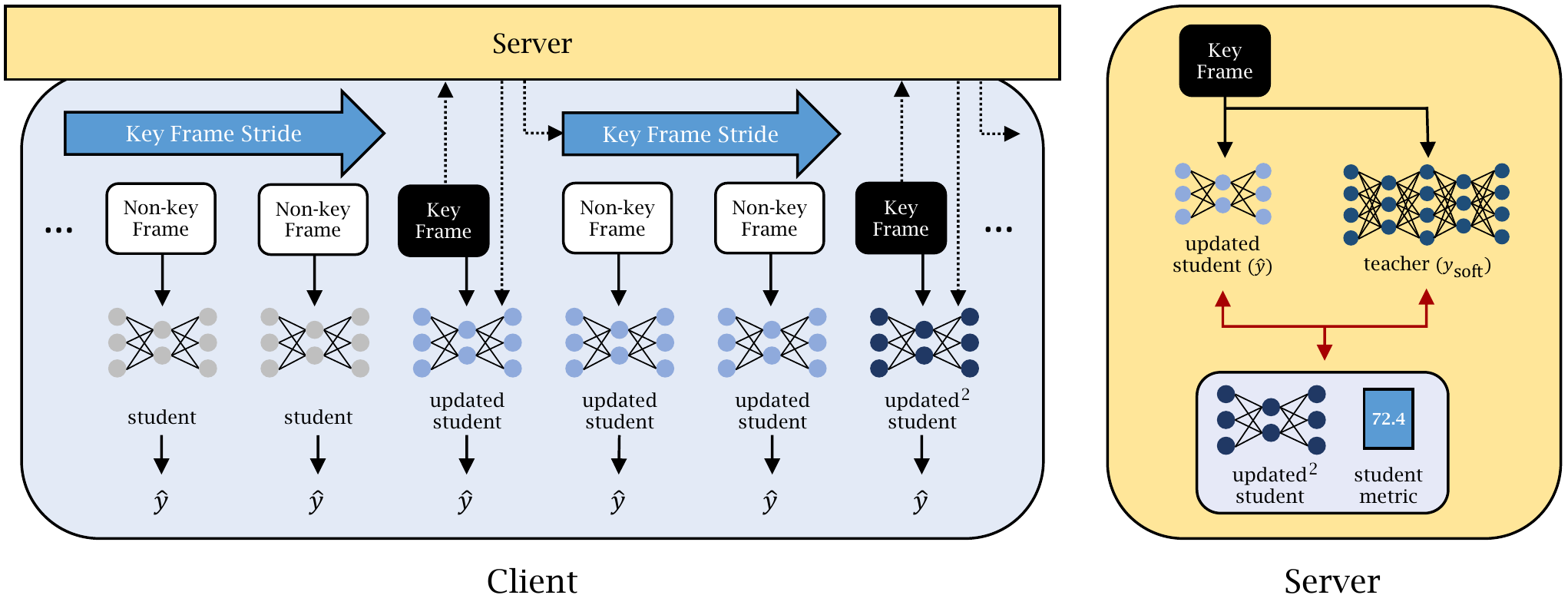}
  \caption{High-level view of ShadowTutor. Dotted arrows denote network communication, while other arrows denote data transfers within a device. Each frame is processed one by one sequentially. Non-key frame inferences are done on-device with a small student network. Sparse key frames are used to update the student's weights via knowledge distillation from a teacher.}
  \label{fig:overview}
  \Description{High-level view of ShadowTutor}
\end{figure*}



Figure \ref{fig:overview} shows an overview of ShadowTutor.
All frames are processed sequentially in temporal order.
Determined by the key frame scheduling policy, if the current frame is a non-key frame, its inference is handled entirely on-device by a lightweight student model.
If the current frame is a key frame, it is sent to the server.
Using the key frame, the server updates the a copy of the student's weights via knowledge distillation from the teacher, and returns the updated weights to the client.
Network communication occurs only at sparse key frames, allowing ShadowTutor to reduce the amount of network traffic drastically.

\subsection{System Components} \label{sec:system-components}

Stripping away the distributed nature from ShadowTutor, there are five essential components that comprise the system: video data, teacher inference, student inference, student training, and key frame striding.

\subsubsection{Video data}

Video data reside in the mobile device, and the majority of the frames do not leave its home.
The video can either be in an internal storage or fetched from the camera in real-time.
Since each frame is traversed by the system in strict temporal order without look-back, frames can be readily discarded once they are processed.

\subsubsection{Teacher inference}

By \textit{teacher}, we refer to a heavy but high-quality neural network deployed to the server.
Since ShadowTutor is an inference system, labels for the input data do not exist.
Thus, it is important to employ a sufficiently good teacher model that can generalize well to unseen video scenes.
Also, the teacher should be pre-trained on a dataset relevant to the task.
Note that teacher inference is done only for the key frames transmitted to the server.

\subsubsection{Student inference}

By \textit{student}, we refer to a lightweight neural network designed to run on the mobile device.
The student model need not be deployed to the mobile device when ShadowTutor is installed, because the server can simply supply the student weights when the system starts.
The student should also be pre-trained on relevant data, most likely the data used to pre-train the teacher.
Pre-training can be expensive, but it is a one-time cost.

For non-key frames, student inference is the only operation done on them.
Hence, student inference latency determines the steady state throughput of ShadowTutor.

\subsubsection{Student training}

\begin{algorithm}[t]
\caption{Student Training}
\label{alg:student-training}
\DontPrintSemicolon
\SetKwData{METRICTHRESHOLD}{THRESHOLD} \SetKwData{MAXUPDATES}{MAX\_UPDATES}
\SetKwFunction{ComputeLoss}{ComputeLoss} \SetKwFunction{ComputeMetric}{ComputeMetric} \SetKwFunction{OptimStep}{OptimStep} \SetKwFunction{Forward}{Forward} \SetKwFunction{PartialBackward}{PartialBackward} \SetKwFunction{Train}{Train}
\SetKwProg{Fn}{Function}{:}{}
\let\oldnl\nl
\newcommand{\nonl}{\renewcommand{\nl}{\let\nl\oldnl}}

\KwIn{Student model $student$, key frame $frame$, pseudo-label $label$}
\KwOut{Trained student model and corresponding metric}

\BlankLine
\nonl \Fn{\Train{$student,~frame,~label$}}{
  $prediction \gets \Forward{student,~frame}$\;
  $best\_metric \gets \ComputeMetric{prediction,~label}$\;
  $best\_student \gets student$\;
  \If{$best\_metric < \METRICTHRESHOLD$}{ \label{alg:line-skip-distillation}
    \For{$i \gets 1$ \KwTo \MAXUPDATES}{
      $loss \gets \ComputeLoss{prediction,~label}$\;
      $gradients \gets \PartialBackward{loss}$\;
      $student \gets \OptimStep{student,~gradients}$\;
      $prediction \gets \Forward{student,~frame}$\;
      $metric \gets \ComputeMetric{prediction,~label}$\;
      \If{$metric > best\_metric$}{
        $best\_metric \gets metric$\;
        $best\_student \gets student$\;
      }
      \If{$metric > \METRICTHRESHOLD$}{
        \textbf{break}
      }
    }
  }
  \Return $(best\_student,~best\_metric)$
}
 
\end{algorithm}

With the teacher's inference result, deemed the pseudo-label, the student model is trained.
Since the student model is very small, it lacks generalization power to perform well on any video stream it encounters.
However, in our setting, there is no need for generalization; we only need to do well for the video at hand.
Thus, by training the student with sparse key frames, we overfit the student model to the current video stream.

Algorithm \ref{alg:student-training} shows the process of student training.
Optimization steps (most simply gradient descent) for the student are taken until either the performance metric (e.g. accuracy, mean IoU) exceeds \texttt{THRESHOLD}, or the number of training steps reach \texttt{MAX\_UPDATES}.
The best performing student and the corresponding metric is returned.
\texttt{THRESHOLD} and \texttt{MAX\_UPDATES} are algorithmic parameters.
The former denotes an acceptable level of student performance, and the latter the maximum number of optimization steps.
Especially, the higher the \texttt{THRESHOLD}, the more the expected number of optimization steps, since it becomes harder for the student to break out of the training loop.
Thus, increasing either of the two potentially increases student performance, but the system's throughput decreases.
Moreover, we perform \textit{partial distillation}, as denoted by the function \texttt{PartialBackward}.
We discuss this in greater detail in section \ref{sec:partial-kd}.

We emphasize that this process be distinguished from student pre-training; student training is done repetitively during system runtime, and student pre-training is done exactly once when the system is first organized.

\subsubsection{Key frame striding}

\begin{algorithm}[t]
\caption{Compute Next Stride}
\label{alg:key-striding}
\DontPrintSemicolon

\SetKwData{METRICTHRESHOLD}{THRESHOLD} \SetKwData{MINSTRIDE}{MIN\_STRIDE} \SetKwData{MAXSTRIDE}{MAX\_STRIDE}
\SetKwFunction{Normalize}{Normalize} \SetKwFunction{Clamp}{Clamp} \SetKwFunction{NextStride}{NextStride}
\SetKwProg{Fn}{Function}{:}{}
\let\oldnl\nl
\newcommand{\nonl}{\renewcommand{\nl}{\let\nl\oldnl}}

\KwIn{Current stride $stride$, student performance metric $metric \in [0,1]$}
\KwOut{Next stride}

\BlankLine
\nonl \Fn{\NextStride{$stride,~metric$}}{
  \eIf{$metric < \METRICTHRESHOLD$}{
    $ratio \gets metric/\METRICTHRESHOLD$ \; \label{alg:left}
  }{
    $ratio \gets (metric - 2*\METRICTHRESHOLD + 1)/(1 - \METRICTHRESHOLD)$ \; \label{alg:right}
  }
  $stride \gets ratio * stride$ \;
  $stride \gets \Clamp{stride,~\MINSTRIDE,~\MAXSTRIDE}$ \;
  \Return $stride$
}

\end{algorithm}

After training the student on the current key frame, the distance to the next key frame must be determined.
It is logical to change the stride, which defines how often to run student training, based on the student's current performance.
However, existing literature regarding key frame striding provides no suitable solution since they are either not adaptive or simplistic (fixed stride \cite{DeepFeatureFlow}, exponential back-off \cite{OD}), or overly complex (clockwork networks \cite{ClockWork}, LSTM \cite{OD++}) for mobile devices.
Thus, we aim design a key frame scheduling algorithm that is sufficiently simple but still adaptive, which is presented in \ref{alg:key-striding}.
Here, the $ratio$ of the next stride to the current one is first determined.
Line \ref{alg:left} is a linear function of $metric$ that connects points $(0,0)$ and $(\texttt{THRESHOLD}, 1)$, and line \ref{alg:right} another linear function connecting $(\texttt{THRESHOLD}, 1)$ and $(1,2)$.
Hence, if the student performs better than \texttt{THRESHOLD}, the distance to the next key frame is elongated, and otherwise shortened.
Finally, to prevent the stride from vanishing or diverging, we clamp the stride with \texttt{MIN\_STRIDE} and \texttt{MAX\_STRIDE}.
These two parameters must be selected based on the data at hand.
For example, if the video has very low FPS and changes quickly even in a few frames, it would be better to choose small numbers for them.
Also, too small values will lead to increased network traffic and deteriorate throughput, while too large values will lead to low student performance.
We analytically model network traffic and throughput in terms of these algorithm parameters in section \ref{sec:throughput-bounds}.

\subsection{Partial Knowledge Distillation} \label{sec:partial-kd}

Instead of changing every parameter of the student, which we call \textit{full} distillation, we suggest to freeze the \emph{front part} of the network, and only train the rest.
This technique is beneficial in terms of latency, memory consumption, and network traffic.
Moreover, it is potentially so for student performance (accuracy).

The reduction in latency and memory consumption is immediate.
Consider the process of back propagation.
Gradients with respect to trainable (not frozen) parameters are computed via chain rule from the loss to the leaf of the computation graph.
Thus, if the front part of a neural network is entirely frozen, gradient computation can stop in the middle of the network.
This leads to decreases in the amount of computation and latency.
Also, memory need not be allocated for the gradients of frozen parameters.

The reduction in network traffic is also evident.
Since both the server and the client agree that only a subset of the student's weights will be adapted, it suffices to communicate only the weights that changed.
Sending the student weights already incurs little overhead because the student is small by design.
Still, partial distillation reduces this overhead further.

The claim of better student performance is rather subtle.
Most neural networks are consisted of a front-end feature extractor, which transforms the input data to a more useful representation, and the back-end network, which maps the representation to a form suitable for the downstream task. 
Thus, given only a limited number of training steps, jointly training both the feature extractor and the back-end network will be costly and unstable.
This is because the front-end network will emit different features every distillation step, and the back-end network must constantly adapt to the change in order to correctly map the feature to the task representation.
Thus, adapting only the weights of the back-end network with a fixed feature distribution can be more stable.
In a sense, with only a small budget for exploration, it is better to invest on exploitation.
Further, drawing reason from existing literature, it is a common practice to freeze the front part of a pre-trained network even for transfer learning \cite{TransferLearning}.
Transfer learning is far more challenging compared with our setting, because it aims to adapt a pre-trained network towards a different task, whereas we stay on the same task.

We empirically validate these claims in section \ref{sec:evaluation} through a series of comparisons with full distillation.

\subsection{ShadowTutor} \label{sec:shadow-tutor}

\begin{algorithm}[t]
\caption{ShadowTutor-Server}
\label{alg:ShadowTutor-server}
\DontPrintSemicolon
\SetKwData{MINSTRIDE}{MIN\_STRIDE}
\SetKwFunction{Recv}{FromClient} \SetKwFunction{Send}{ToClient} \SetKwFunction{UpdatedPart}{UpdatedPart} \SetKwFunction{Forward}{Forward} \SetKwFunction{Train}{Train} 
\KwIn{Student model $student$, teacher model $teacher$}
\BlankLine

 \Send{student}\;
 \While{forever}{
   \Recv{frame}\;
   $label \gets \Forward{teacher,~frame}$\;
   $student,~metric \gets \Train{student,~frame,~label}$\;
   $\Send{\UpdatedPart{student},~metric}$\;
 }
 
\end{algorithm}

\begin{algorithm}[t]
\caption{ShadowTutor-Client}
\label{alg:ShadowTutor-client}
\DontPrintSemicolon
\SetKwData{MINSTRIDE}{MIN\_STRIDE}
\SetKwFunction{Recv}{FromServer} \SetKwFunction{AsyncRecv}{FromServerAsync} \SetKwFunction{AsyncSend}{ToServerAsync} \SetKwFunction{UpdatedPart}{UpdatedPart} \SetKwFunction{Forward}{Forward} \SetKwFunction{Completed}{Completed} \SetKwFunction{WaitUntilComplete}{WaitUntilComplete} \SetKwFunction{ComputeNextStride}{NextStride} \SetKwFunction{ApplyUpdate}{ApplyUpdate}
\KwIn{Target video stream $video$}
\KwOut{Student predictions}
\BlankLine

 $stride \gets \MINSTRIDE$\;
 $step \gets stride$ \tcp*{First frame as key frame}
 $updated \gets \texttt{true}$ \tcp*{Whether student recv is done}
 $\Recv{student}$\;
 \ForEach{$frame$ \upshape{in} $video$}{
   \If(\tcp*[f]{Key frame}){$step = stride$}{
     $\AsyncSend{frame}$\; \label{alg:async-send}
     $async\_recv \gets \AsyncRecv{student\_diff,~metric}$\; \label{alg:async-recv}
     $step \gets 0$\;
     $updated \gets \texttt{false}$
   }
   $prediction \gets \Forward{student,~frame}$\;
   $step \gets step + 1$\;
   \If{\upshape{not} $updated$}{
     \If{$step = \MINSTRIDE$}{
       $\WaitUntilComplete{async\_recv}$\;
     }
     \If{$\Completed{async\_recv}$}{
       $student \gets \ApplyUpdate{student,~student\_diff}$\;
       $stride \gets \ComputeNextStride{stride,~metric}$\;
       $updated \gets \texttt{true}$
     }
   }
 }
 
\end{algorithm}

Now, we aggregate the system components into ShadowTutor.
Algorithm \ref{alg:ShadowTutor-server} and \ref{alg:ShadowTutor-client} shows the runtime operations of the server and the mobile device (client) respectively.
ShadowTutor offloads teacher inference and student training to the server.
Needless to say, teacher inference is an unacceptable workload for mobile devices, so offloading is a natural choice.
Student training on mobiles devices is feasible, but it still incurs significant delay.
Also, student training blocks student inference; the mobile device cannot proceed with non-key frame inference during student training.
This causes large fluctuations in the system's throughput.
On the other hand, for the server, training the student only adds marginal overhead since the student only has a small number of parameters.

ShadowTutor reduces network traffic considerably.
This is because network communications between the server and the client only occurs at key frames, as opposed to naive offloading, which requires communication for every frame.
In addition, thanks to partial distillation, the data size of each network communication is cut down further.
Thus, ShadowTutor reduces network traffic in terms of the number of network transmissions and the number of packets in each transmission.

Importantly, offloading student training reveals an opportunity for \textit{asynchronous} inference.
As shown in line \ref{alg:async-send} and \ref{alg:async-recv} of algorithm \ref{alg:ShadowTutor-client}, the mobile device sends key frames and receives the updated student parameters in a non-blocking fashion.
That is, the mobile device sends the key frame to the server, and without waiting for the updated parameters, proceeds on to the next (non-key) frame.
This approach further exploits the temporal coherence between frames by assuming that even after a key frame, the student weights are still usable.
The updated parameters are awaited for a maximum \texttt{MIN\_STRIDE} steps, since the next key frame stride may become \texttt{MIN\_STRIDE}.
This is the key mechanism that allows ShadowTutor to be robust to adverse network conditions.
That is, ShadowTutor can mitigate delays in network transfer by just proceeding further into future frames.

\subsection{Network Traffic and Throughput Bounds} \label{sec:throughput-bounds}

In this section, we model ShadowTutor's network traffic (amount of data transferred per unit time) and throughput (number of frames processed per unit time) and derive formulae for their lower and upper bounds.
We use the notations defined in table \ref{tab:notations}.
Note that $t_{net}$ and $s_{net}$ regard the transmission of one key frame and the corresponding updated student parameters.

\begin{table}[t]
  \centering
  
  \caption{Notations used for ShadowTutor. Those in the first block are identified after system execution, and the second based on system component decisions.}
  \label{tab:notations}
  
  \begin{tabular}{ll}
  \toprule
  Symbol  & Definition \\
  \midrule
  $n$    & \textbf{n}umber of frames processed \\
  $d$    & number of \textbf{d}istillation steps taken \\
  $k$    & number of \textbf{k}ey frames \\
  \midrule
  $t_{si}$     & latency of \textbf{s}tudent \textbf{i}nference \\
  $t_{sd}$     & latency of one \textbf{s}tudent \textbf{d}istillation step \\
  $t_{ti}$     & latency of \textbf{t}eacher \textbf{i}nference \\
  $t_{net}$    & \textbf{net}work latency associated with one key frame \\
  $s_{net}$    & \textbf{net}worked data \textbf{s}ize associated with one key frame \\
  \bottomrule
  \end{tabular}
\end{table}

Both network traffic and throughput rely on the system's total execution time, which we first model.
ShadowTutor is designed to perform asynchronous inference for at most \texttt{MIN\_STRIDE} many frames after a key frame.
However, a mobile device may either be able to execute student inference and network operations entirely in parallel, or it may not support any form of concurrency.
Therefore, $t_c$, the execution time of \texttt{MIN\_STRIDE} frames after a key frame, is within the following bounds:
\begin{equation} \label{eq:t-c}
\begin{split}
    t_c & \geq \max(\texttt{MIN\_STRIDE} \times t_{si}, t_{net} + t_{ti}) \\
    t_c & \leq \texttt{MIN\_STRIDE} \times t_{si} + t_{net} + t_{ti}
\end{split}
\end{equation}
Then, with $t_c$, the total execution time for processing $n$ frames can be modelled as follows:
\begin{equation} \label{eq:t-tot}
    t_{tot} = (n - k \times \texttt{MIN\_STRIDE}) t_{si} + d t_{sd} + k t_c.
\end{equation}

Now, we obtain a general formula for network traffic by dividing the total size of data transfer by the total execution time:
\begin{equation} \label{eq:nt-general}
    \frac{k s_{net}}{t_{tot}} = \frac{k s_{net}}{(n - k \times \texttt{MIN\_STRIDE}) t_{si} + d t_{sd} + k t_c}.
\end{equation}

Minimum network traffic is achieved when key frames are least frequent, the execution time for each key frame is longest, and the client completely lacks concurrency.
That is,
\begin{equation} \label{eq:least-keys}
    k = \frac{n}{\texttt{MAX\_STRIDE}},
\end{equation}
\begin{equation} \label{eq:most-distills}
    d = k \times \texttt{MAX\_UPDATES},
\end{equation}
and
\begin{equation} \label{eq:min-concurrency}
    t_c = \texttt{MIN\_STRIDE} \times t_{si} + t_{net} + t_{ti}
\end{equation}
hold.
Thus, from equation \ref{eq:nt-general}, the network traffic lower bound is:
\begin{equation} \label{eq:nt-lower}
    \frac{s_{net}}{\texttt{MAX\_STRIDE} \times t_{si} + \texttt{MAX\_UPDATES} \times t_{sd} + t_{ti} + t_{net}}.
\end{equation}
On the other hand, network traffic is maximum when key frames are as frequent as possible, the execution time for each key frame is shortest, and the client is capable of handling student inference and network operations entirely in parallel.
That is,
\begin{equation} \label{eq:most-keys}
    k = \frac{n}{\texttt{MIN\_STRIDE}},
\end{equation}
\begin{equation} \label{eq:least-distills}
    d = 0,
\end{equation}
and
\begin{equation} \label{eq:max-concurrency}
    t_c = \max(\texttt{MIN\_STRIDE} \times t_{si}, t_{net} + t_{ti})
\end{equation}
hold.
Especially, equation \ref{eq:least-distills} holds because distillation can be entirely skipped based on the student's initial metric (see line \ref{alg:line-skip-distillation} in algorithm \ref{alg:student-training}).
Again, from equation \ref{eq:nt-general}, the network traffic upper bound is:
\begin{equation} \label{eq:nt-upper}
    \frac{s_{net}}{\max(\texttt{MIN\_STRIDE} \times t_{si}, t_{net} + t_{ti})}.
\end{equation}

We can also obtain a general formula for throughput by dividing the number of processed frames by the execution time:
\begin{equation} \label{eq:tp-general}
    \frac{n}{t_{tot}} = \frac{n}{(n - k \times \texttt{MIN\_STRIDE}) t_{si} + d t_{sd} + k t_c}.
\end{equation}

The throughput lower bound is achieved when the total execution time is the longest.
In such case, equations \ref{eq:most-keys}, \ref{eq:most-distills}, and \ref{eq:min-concurrency} hold.
Thus, from equation \ref{eq:tp-general}, the throughput lower bound is
\begin{equation} \label{eq:tp-lower}
    \frac{\texttt{MIN\_STRIDE}}{\texttt{MIN\_STRIDE} \times t_{si} + \texttt{MAX\_UPDATES} \times t_{sd} + t_{ti} + t_{net}}.
\end{equation}
On the other hand, the throughput upper bound is achieved when the total execution time is the shortest.
In that case, equations \ref{eq:least-keys}, \ref{eq:least-distills}, and \ref{eq:max-concurrency} hold.
Again, from equation \ref{eq:tp-general}, the throughput upper bound is
\begin{equation} \label{eq:tp-upper}
    \frac{\texttt{MAX\_STRIDE}}{(\texttt{MAX\_STRIDE} - \texttt{MIN\_STRIDE}) t_{si} + \max(\texttt{MIN\_STRIDE} t_{si}, t_{net} + t_{ti})}.
\end{equation}

Notice that in all lower and upper bound formulae, only algorithm parameters, latency measurements, and data size remain.
Thus, ShadowTutor allows the estimation of the system's network bandwidth requirement and throughput prior to actually implementing and running the entire system.
We use these bounds to determine algorithm parameters in section \ref{sec:algorithm-param-decisions}.

\section{ShadowTutor for Video Semantic Segmentation} \label{sec:ShadowTutor-semseg}






\subsection{Experiment setup} \label{sec:setup}

As the server, we use a desktop computer equipped with an AMD Ryzen 7 3700X CPU, one NVIDIA RTX 2080ti GPU, and 32 GB memory.
As the client, we use the NVIDIA Jetson Nano embedded board \cite{Jetson}, equipped with a quad-core ARM A57 CPU, a 128-core Maxwell GPU, and 4 GB memory.
Jetson Nano can deliver up to 472 GFLOPS for 32-bit floating points, which is not an unrealistic number for modern mobile devices.
For example, Google Pixel 4's Qualcomm Snapdragon 855 can deliver up to 954.7 GFLOPS (32-bit) with its built-in Adreno 640 GPU \cite{Adreno640}.
As to network configurations, we limit both uplink and downlink bandwidth to 80 Mbps, assuming strong Wi-Fi connection.
We implement the system with OpenMPI \cite{OpenMPI}, PyTorch \cite{PyTorch}, and Detectron2 \cite{Detectron2}.

\subsection{System Component Decisions} \label{sec:system-component-decisions}

We target videos with 25--30 FPS from the Long Video Segmentation (LVS) dataset \cite{OD} \footnote{Obtained from \url{https://olimar.stanford.edu/hdd/lvsdataset/}.}.
We use high resolution (720p HD) videos in order to pressure the overall load of the system.
The LVS dataset is labeled with 8 actively moving object classes (person, bicycle, automobile, bird, dog, horse, elephant, and giraffe), making accurate segmentation challenging because no object class remains stationary in the scene.
The movement of the camera is either fixed, moving, or egocentric (shot from a camera attached to a person's head or chest).
Also, the main scenery of each video is one of the three: animals, people, or street.

We choose Mask R-CNN \cite{MaskRCNN} as the teacher neural network, because the LVS dataset was actually labelled by selecting videos that Mask R-CNN performs particularly well on.
The teacher's weights (pre-trained on the Microsoft COCO dataset \cite{COCO}) are adopted from the Detectron2 framework.

\begin{figure}[t]
  
  \subcaptionbox{Operations of a student block\label{fig:student-block}}{
    \centering
    \includegraphics[width=6cm]{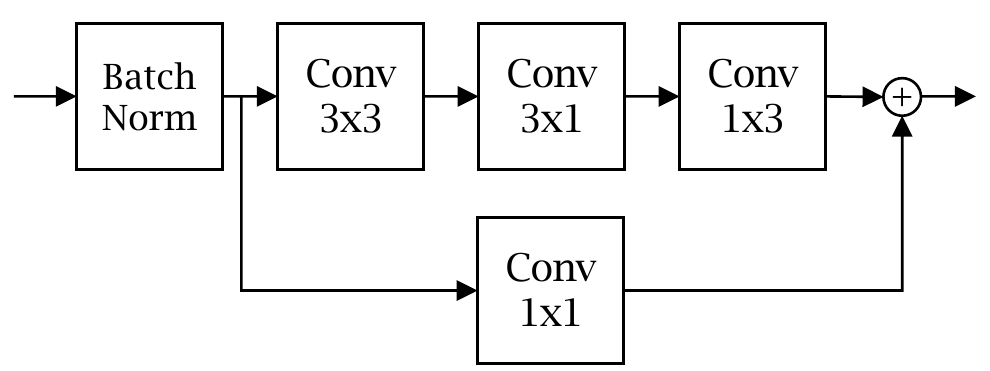}
  }
  \subcaptionbox{Student architecture\label{fig:student-arch}}{
    \centering
    \includegraphics[width=8cm]{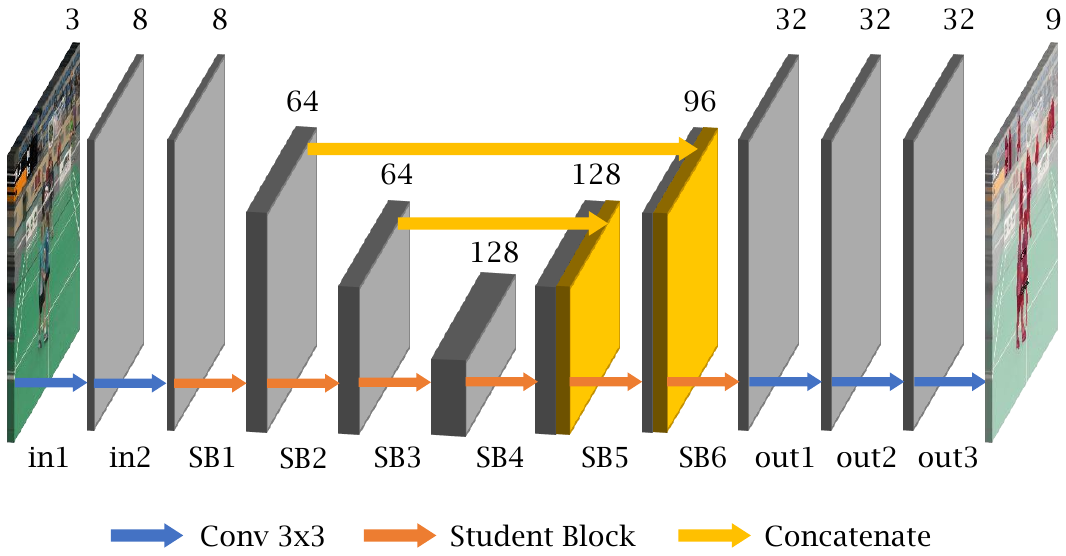}
  }
  
  \caption{The student model is a small fully-convolutional network.}
\end{figure}

The student neural network is a simple fully-convolutional network.
Figure \ref{fig:student-block} shows the operations of a student block, and \ref{fig:student-arch} the layer compositions of the student.
The low resolution feature maps from \texttt{SB2} and \texttt{SB1} are concatenated to the input of \texttt{SB5} and \texttt{SB6}, respectively.
We pre-trained the student on COCO for 30 epochs with the Adam \cite{Adam} optimizer.
As to partial distillation, we freeze the student from the first layer to \texttt{SB4}, only computing gradients until \texttt{SB5}.
This way, the trained parameters amount to 21.4\% of all parameters.


  

Counting the raw number of parameters, the teacher has 44.34 million parameters, whereas the student has 0.48 million.
Thus, the teacher is 100x larger than the student.
However, as we will see in section \ref{sec:eval-accuracy}, the student can approach the teacher's performance through shadow education.

Model-level optimizations such as using more efficient operations (as in MobileNet \cite{MobileNetV2}) or performing quantization or pruning on weights \cite{DeepComp} can be applied to the student.
Further, teacher inference can be accelerated with inference optimizations such as TensorRT \cite{TensorRT}.
However, although these techniques may improve throughput and decrease network traffic, we exclude them in our study.
This is because the effect of such techniques differ widely based on implementation, platform, and hardware.

Finally, we decide upon the actual process of knowledge distillation.
As pointed out by the LVS dataset paper, videos in the dataset have an excess amount of background class pixels.
Due to this class imbalance, training the student with vanilla cross entropy may bias the student towards predicting every pixel as background.
Thus, we directly adopt their loss weighting approach, which scales the cross-entropy loss of pixels near and within non-background objects by a factor of 5.
Partial knowledge distillation to the student parameters is done with the Adam optimizer with a learning rate of 0.01.

\subsection{Algorithm Parameter Decisions} \label{sec:algorithm-param-decisions}

We first determine \texttt{THRESHOLD} using the performance of existing video semantic segmentation models.
Since the LVS dataset has no official leaderboard available, we consider a similar dataset called Cityscapes \cite{Cityscapes}.
Since the state-of-the-art approach for Cityscapes records an mIOU of 0.845 at the time of research, we set \texttt{THRESHOLD} to 0.8.

As to \texttt{MIN\_STRIDE} and \texttt{MAX\_STRIDE}, we consider the FPS of the videos.
We have selected videos with an FPS of 25--30.
Following this, we set \texttt{MIN\_STRIDE} to 8 and \texttt{MAX\_STRIDE} to 64.
That is, within 8 frames, we postulate that the objects in the scene will have hardly changed, and re-training within 8 frames would be meaningless.
On the other hand, after 64 frames or 2.5 seconds in a video clip, we assume that the objects will have moved significantly, and the student at least needs to be tested.

Finally, we determine \texttt{MAX\_UPDATES} using the throughput bounds derived in equation \ref{eq:tp-lower} and \ref{eq:tp-upper}.
Using the notations defined in table \ref{tab:notations}, our experiment setting gives $t_{si} = 0.143$, $t_{sd} = 0.013$, $t_{ti} = 0.044$, and $t_{net} = 0.303$ in seconds.
Then, equation \ref{eq:tp-upper} yields a maximum throughput of 6.99 FPS.
In order to keep the difference between the theoretical maximum and minimum throughput within 2 FPS, we find the largest \texttt{MAX\_UPDATES} value that gives a throughput lower bound larger than 5 FPS, which is 8.

\section{Evaluation} \label{sec:evaluation}

Using the configurations from section \ref{sec:ShadowTutor-semseg}, we show ShadowTutor's effectiveness.
Specifically, we investigate the advantages of ShadowTutor in terms of throughput, network traffic, and accuracy in sections \ref{sec:eval-throughput}, \ref{sec:eval-network-traffic}, and \ref{sec:eval-accuracy}, respectively.
Then, we push ShadowTutor to the limits in terms of network bandwidth in section \ref{sec:eval-network-robust} and temporal coherence in section \ref{sec:eval-temporal-coherence}.

ShadowTutor is mainly compared with naive offloading (sending every frame to the server and retrieving inference results).
Also, we compare partial distillation and full distillation to justify the design of ShadowTutor.
All experiments are performed on the first 5000 frames of each video stream (about 200 seconds).
Every ShadowTutor experiment, whether partial or full distillation, begin from the same pre-trained student checkpoint.

In or experiments, we only employ one type of teacher model: the Mask R-CNN.
This is because the student, who learns from the teacher, is only interested in the final output of the teacher, regardless of all the intermediate operations.

\subsection{Throughput} \label{sec:eval-throughput}

\begin{table}[t]
  \centering
    
  \caption{Execution time and mean number of distillation steps}
  \label{tab:distill-stats}
  
  \begin{tabular}{lll}
  \toprule
  Distillation          & Partial   & Full\\
  \midrule
  One step (ms)         & 13        & 18\\
  Mean \# of steps      & 3.83      & 4.44\\
  \bottomrule
  \end{tabular}
\end{table}

\begin{table}[t]
  \centering
    
  \caption{Frames processed per second (FPS) and execution time (s) in parenthesis}
  \label{tab:fps-exectime}
  
  \begin{tabular}{lllll}
  \toprule
  Camera     & Scene   & Partial              & Full   & Naive\\
  \midrule
  fixed      & animals & 6.55(762.5)          & 6.21(804.5) & 2.09(2391.3)\\
  fixed      & people  & \textbf{6.60(757.4)} & 6.43(777.0) & 2.09(2391.3)\\
  fixed      & street  & 6.50(768.8)          & 5.95(840.5) & 2.09(2391.3)\\
  moving     & animals & 6.57(760.5)          & 6.27(796.5) & 2.09(2391.3)\\
  moving     & people  & \textbf{6.59(758.5)} & 6.36(785.8) & 2.09(2391.3)\\
  moving     & street  & 6.41(780.2)          & 5.55(901.0) & 2.09(2391.3)\\
  egocentric & people  & 6.57(760.5)          & 5.89(848.5) & 2.09(2391.3)\\
  \midrule
  \multicolumn{2}{c}{average} & \textbf{6.54(764.1)} & 6.08(822.0) & 2.09(2391.3)\\
  \bottomrule
  \end{tabular}
\end{table}

Table \ref{tab:distill-stats} summarizes the latency of one distillation step (ms) and the mean number of distillation steps for partial and full distillation.
Partial distillation reduces both the latency and the number of distills, contributing to the throughput of the entire system.
This result empirically supports the claim that since the number of training steps is limited, it is quicker to exploit a fixed distribution of features than to explore for better ones.

Table \ref{tab:fps-exectime} lists the actual throughput of the system (frames processed per second) and the total execution time.
As expected, partial distillation outperforms full distillation in every category.
Moreover, ShadowTutor shows an improvement greater than 3x over naive offloading.
This is especially because ShadowTutor only communicates with the server on key frames, and thus drastically reduces the latency for networking.

\subsection{Network Traffic} \label{sec:eval-network-traffic}

\begin{table}[t]
  \centering
    
  \caption{Data transmitted on each key frame (MB).}
  \label{tab:network-amount}
  
  \begin{tabular}{llll}
  \toprule
  Direction & Partial        & Full  & Naive\\
  \midrule
  To Server & 2.637          & 2.637 & 2.637\\
  To Client & \textbf{0.395} & 1.846 & 0.879\\
  \midrule
  Total     & \textbf{3.032} & 4.483 & 3.516\\
  \bottomrule
  \end{tabular}
\end{table}

\begin{table}[t]
  \centering
  
  \caption{Key frames ratio (\%) and network traffic (Mbps)}
  \label{tab:key-frame-ratio}
  
  \begin{tabular}{lllllll}
  \toprule
  \multirow{2}{*}{Camera} & \multirow{2}{*}{Scene} & \multicolumn{3}{c}{Key frame ratio} & \multicolumn{2}{c}{Network traffic}\\
  \cmidrule(lr){3-5}
  \cmidrule(lr){6-7}
  && Partial & Full & Naive & Partial & Naive\\
  \midrule
  fixed      & animals        & 4.73            & 4.60   & 100.0 & 7.51          & 58.51\\
  fixed      & people         & \textbf{1.96}   & 2.42   & 100.0 & \textbf{3.14} & 58.51\\
  fixed      & street         & 7.78            & 7.43   & 100.0 & 12.27         & 58.51\\
  moving     & animals        & 2.55            & 2.29   & 100.0 & 4.06          & 58.51\\
  moving     & people         & \textbf{3.45}   & 4.12   & 100.0 & 5.51          & 58.51\\
  moving     & street         & 11.70           & 11.48  & 100.0 & 18.19         & 58.51\\
  egocentric & people         & \textbf{5.46}   & 9.75   & 100.0 & 8.70          & 58.51\\
  \midrule
  \multicolumn{2}{c}{average} & \textbf{5.38}   & 6.01   & 100.0 & \textbf{6.19} & 58.51\\
  \bottomrule
  \end{tabular}
\end{table}

We investigate the reduction of network traffic in terms of the amount of data transfer per key frame and the ratio of key frames to all frames.

Table \ref{tab:network-amount} shows the amount of data transfer (in MB) per key frame.
Since it suffices to send only the updated part of the student, partial distillation reduces network traffic compared with full distillation.
Against naive offloading, which sends the teacher prediction to the client, ShadowTutor reduces the amount of data transfer by 13.77\% per key frame, because the size of the student is even smaller than one video frame.

Table \ref{tab:key-frame-ratio} summarizes the proportion of key frames and the actual network traffic in Mbps.
The smaller the key frame proportion, the less frequent the network communication.
Partial distillation generally performs better than full distillation, and strictly better than naive offloading.
Especially, for the fixed-people category, the number of network communications is only 1.96\% compared with the naive offloading scheme, yielding a surprising 98\% reduction.

The effects of the two reductions are multiplicative.
Thus, compared with naive offloading, ShadowTutor reduces the amount of network transfer per frame by 98.3\% at most, and 95.3\% on average.

On the other hand, the reduction in network traffic (amount of networked data per unit time) is coupled with the improvement in throughput, showing a reduction of 89.4\% on average.
We especially note that network traffic has improved even if throughput had a threefold improvement.

Finally, with the current configuration, the network traffic bounds computed with equations \ref{eq:nt-lower} and \ref{eq:nt-upper} are 2.53 Mbps and 21.2 Mbps, respectively.
From table \ref{tab:key-frame-ratio}, all network traffic values obey the bounds, and the bounds are quite tight, proving their usefulness.

\subsection{Accuracy} \label{sec:eval-accuracy}

\begin{table}[t]
  \centering
    
  \caption{Mean IoU of various settings. Wild = pre-trained student on its own, P = partial distillation, F = full distillation, digit (1 or 8) = number of delayed frames before receiving updated student weights.}
  \label{tab:mIoU}
  
  \begin{tabular}{*7l}
  \toprule
  Camera     & Scene   & Wild  & P-1            & P-8   & F-1   & Naive\\
  \midrule
  fixed      & animals & 14.34 & 74.31          & 73.27 & 74.47 & 100.0 \\
  fixed      & people  & 13.91 & 81.69          & 81.39 & 81.36 & 100.0 \\
  fixed      & street  & 17.28 & \textbf{70.26} & 69.01 & 63.60 & 100.0 \\
  moving     & animals & 22.31 & 74.94          & 73.80 & 75.21 & 100.0 \\
  moving     & people  & 17.62 & 74.82          & 74.06 & 75.55 & 100.0 \\
  moving     & street  & 18.65 & \textbf{60.48} & 58.61 & 52.94 & 100.0 \\
  egocentric & people  & 14.80 & \textbf{70.42} & 68.87 & 61.41 & 100.0 \\
  \midrule
  \multicolumn{2}{c}{average} & 16.99 & \textbf{72.42} & 71.29 & 69.22 & 100.0\\
  \bottomrule
  \end{tabular}
\end{table}

Table \ref{tab:mIoU} shows the mean Intersection over Union (mIoU) of various experiment settings.
The mIoU of \emph{every} frame (key and non-key frames) is averaged to show that the student can leverage temporal coherence to accurately perform inference on non-key frames.
Note that all accuracy values are evaluated against the teacher (Mask R-CNN) output, which is why the naive approach always achieves perfect accuracy.
However, we emphasize that the LVS dataset has been labelled with the Mask R-CNN.
Thus, in our case, we are measuring the accuracy against the label in effect.

First, to show the need for shadow education, we run the pre-trained student on every frame without any supervision from the teacher (denoted as Wild).
As expected, its accuracy suffers greatly, approaching the accuracy of random guessing.
This is because the student is too small to generalize to all kinds of scenes.

Next, we show the effect of shadow education.
Recall that the mobile device receives the updated weights in a non-blocking fashion, mitigating the effect of delays in network transfer.
Thus, we measure accuracy when there is the least delay (1 frame, P-1) and the most (8 frames, P-8).
ShadowTutor approaches the accuracy of the teacher with a student 100x smaller, proving the effectiveness of knowledge distillation.
Moreover, asynchronous inference hardly hurts accuracy, showing that slightly outdated weights are still useful due to temporal coherence.

Lastly, we compare partial distillation (P-1) and full distillation (F-1).
Overall, partial distillation is more accurate.
When partial distillation is better, it outperforms full distillation significantly, thereby providing an overall stable level of accuracy.

\subsection{Robustness to Network Conditions} \label{sec:eval-network-robust}

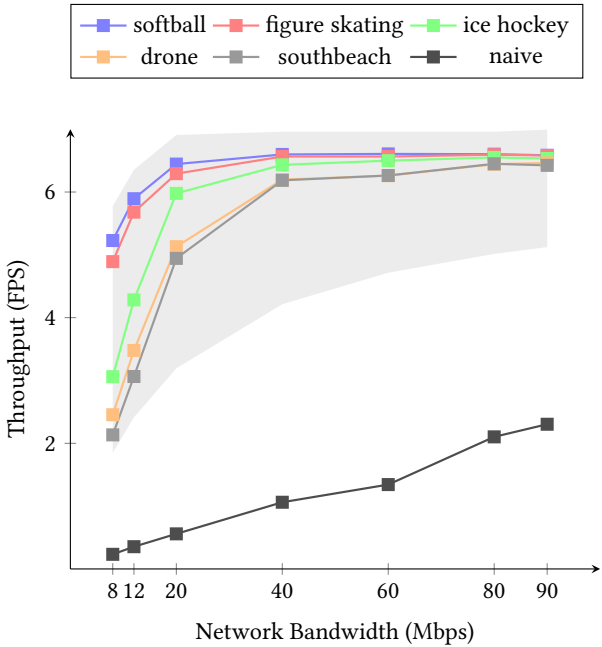
\begin{figure}
\centering
    \begin{adjustbox}{width=0.45\textwidth}
    \begin{tikzpicture}
    \begin{axis}[
        axis lines=middle,
        ymin=0,
        ymax=7,
        xmin=0,
        xmax=100,
        xlabel=Network Bandwidth (Mbps),
        ylabel=Throughput (FPS),
        x label style={at={(axis description cs:0.5,-0.1)},anchor=north},
        y label style={at={(axis description cs:-0.06,.5)},rotate=90,anchor=south},
        legend style={at={(0.0,1.2)},anchor=west,legend columns=3},
        enlargelimits = false,
        xticklabels from table={data/BW-FPS.dat}{BW},xtick=data]+
        
    \addplot[name path=upper,draw=none,forget plot] table [x=BW,y=upper]{data/BW-FPS.dat};
    \addplot[name path=lower,draw=none,forget plot] table [x=BW,y=lower]{data/BW-FPS.dat};
    \addplot[fill=gray!15,forget plot] fill between[of=upper and lower];
    
    \addplot[blue!50,thick,mark=square*] table [y=softball,x=BW]{data/BW-FPS.dat};
    \addlegendentry{softball}
    \addplot[red!50,thick,mark=square*] table [y=figure-skating,x=BW]{data/BW-FPS.dat};
    \addlegendentry{figure skating}
    \addplot[green!50,thick,mark=square*] table [y=ice-hockey-ego,x=BW]{data/BW-FPS.dat};
    \addlegendentry{ice hockey}
    \addplot[orange!50,thick,mark=square*] table [y=drone,x=BW]{data/BW-FPS.dat};
    \addlegendentry{drone}
    \addplot[gray!80,thick,mark=square*] table [y=southbeach,x=BW]{data/BW-FPS.dat};
    \addlegendentry{southbeach}
    \addplot[black!70,thick,mark=square*] table [y=naive,x=BW]{data/BW-FPS.dat};
    \addlegendentry{naive}
    
    \end{axis}
    \end{tikzpicture}
    \end{adjustbox}
\caption{Network bandwidth and system throughput}
\label{fig:network-robust}
\end{figure}


Fluctuations often happen during network communications between the cloud data center and the client.
Thus, in this experiment, we investigate the effect of reduced available network bandwidth.
Specifically, we set the bandwidth of the system to 90, 80, 60, 40, 20, 12, and 8 Mbps, and examine the throughput of the system.

Figure \ref{fig:network-robust} shows the change in throughput against network bandwidth for ShadowTutor and naive offloading.
For ShadowTutor, we selected five video streams with different key frame proportions; softball has the least key frames (1.72\%), and southbeach (street CCTV) has the most (12.4\%).

The throughput of naive offloading decreases immediately in the face of low network bandwidth because it has no mechanism to mitigate the increase in network latency.
On the contrary, the throughput of ShadowTutor remains remarkably stable until 40 Mbps, which is half of the original bandwidth.
For videos that have a small proportion of key frames, throughput is retained even until 20 Mbps, since network latency takes up only a small fraction among all latency components.
Videos with more key frames lose throughput more quickly, but only by 3x even if the network bandwidth is 10x narrower.

The region colored in gray represent the throughput bounds computed with equations \ref{eq:tp-lower} and \ref{eq:tp-upper}.
All throughput values obey the bounds.
Especially, for low bandwidth settings, network latency dominates among all latency components, reducing the variation in throughput brought about by the degree of concurrency supported by the mobile device.

ShadowTutor's robustness to the reduction in network bandwidth comes from asynchronous inference.
In effect, as long as the network latency is shorter than the inference latency of \texttt{MIN\_STRIDE} many frames, ShadowTutor can hide the network latency almost completely.
However, when the network latency is longer, the reduction in network bandwidth begins to take a more direct impact to the system's throughput since asynchronous inference can no longer serve as a buffer.

\subsection{Feasibility of Real-Time Inference} \label{sec:eval-temporal-coherence}

\begin{table}[t]
  \centering
    
  \caption{Mean IoU and key frame ratio for 7 FPS videos. Digit (1 or 8) = number of frame delays before receiving updated student weights. Key frame proportion is in \%.}
  \label{tab:temporal-coherence}
  
  \begin{tabular}{*5l}
  \toprule
  Camera     & Scene   & Partial-1   & Partial-8   & Key frame     \\
  \midrule
  fixed      & animals & 62.72 & 61.86 & 6.59 \\
  fixed      & people  & 80.44 & 80.08 & 1.97 \\
  fixed      & street  & 63.78 & 62.51 & 8.9  \\
  moving     & animals & 68.63 & 66.78 & 4.84 \\
  moving     & people  & 73.66 & 72.91 & 4.15 \\
  moving     & street  & 48.92 & 46.99 & 12.34\\
  egocentric & people  & 67.57 & 66.09 & 5.44 \\
  \midrule
  \multicolumn{2}{c}{average} & 66.53 & 65.31 & 6.32\\
  \bottomrule
  \end{tabular}
\end{table}

Finding promise from 25--30 FPS videos, we test ShadowTutor with videos with less temporal coherence.
Specifically, for every video, we re-sample the frames such that all videos have an FPS of 7.
Thus, by matching the input video's frame rate with ShadowTutor's throughput, we simulate the \emph{real-time inference} of frames fetched from the mobile device's camera.

Table \ref{tab:temporal-coherence} shows the mean IoU and key frame proportion of the first 5000 frames of the re-sampled videos.
Surprisingly, even if the time distance between adjacent frames is elongated by four times, ShadowTutor  yields an average accuracy drop of less than 6\%p and key frame proportion increase of less than 1\%p.
Therefore, this shows the feasibility of applying ShadowTutor to real-time inference applications.
With optimized students, e.g. those that utilize weight quantization/pruning or employ more efficient operations, ShadowTutor will be able to handle higher frame rate videos in real-time, and its accuracy will increase further due to stronger temporal coherence.


  
  






\section{Related Work} \label{sec:related-work}


We briefly survey works related to extending or replacing parts of ShadowTutor.

The concept of knowledge distillation has been applied to computer vision tasks widely.
Chen et al. \cite{ObjectDetKD} applied knowledge distillation to creating efficient object detection models for images.
Mullapudi et al. \cite{OD} extended knowledge distillation to videos.
Bajestani et al. \cite{OD++} extends the previous work by designing an LSTM-based key frame selection method and a teacher-bounded loss function.
These works do not consider the context of mobile or distributed computing.
Still, they provide insight into knowledge distillation and the possibility of direct application to ShadowTutor.

We may delve further into the possibility of extending ShadowTutor's knowledge distillation process.
The original knowledge distillation paper \cite{KD} also proposed to distill knowledge from an \emph{ensemble} of different teacher models.
Moreover, data distillation \cite{DataDistillation} proposes to use only a single teacher, but to ensemble its outputs on the same image but with different transformations applied.
Chung et al. \cite{FeatureAdversarialKD} proposes to apply knowledge distillation at the \emph{feature}-level through adversarial training.

\section{Conclusion and Future Work} \label{sec:conclusion}

In this work, we developed ShadowTutor, a distributed video DNN inference framework that encodes the temporal coherence in the video at hand into the parameters of a small student model through intermittent partial knowledge distillation.
Evaluations show its advantages in terms of network traffic, throughput, accuracy, and robustness.
Moreover, ShadowTutor achieves this through a simple and elegant split between the server and the client, and without complex DNN partitioning, server-side scheduling, or model-level optimizations tailored to the experiment devices.

While our work focused on developing a distributed framework for video data, ShadowTutor can also be extended to tasks other than video computer vision.
That is, there are plenty of \textit{sequence data}, i.e. a collection of \textit{data points} that are temporally coherent, that are handled with DNNs.
Examples of such sequence data include speech signals from a single speaker, a sentence that requires translation, or a series of item recommendation requests from a single user.
Thus, by exploiting the temporal coherence embedded in various types of sequence data through intermittent knowledge distillation, ShadowTutor has the potential to be extended to any type of sequence data.
Upgrading the knowledge distillation process, designing appropriate model architectures for the teacher and the student, and resolving the issues that may arise in the process will serve as a good future research direction.


\begin{acks} 
This work was supported by the National Research Foundation of Korea (NRF) grant funded by the Korea government (MSIT) (No. 2020R1A2B5B02001845).
\end{acks}

\bibliographystyle{ACM-Reference-Format}
\bibliography{main}


\end{document}